\PassOptionsToPackage{table, dvipsnames}{xcolor}

\documentclass[sigconf]{acmart}

\usepackage{soul}
\usepackage{enumitem}
\usepackage{xspace}
\usepackage{subcaption}

\newcommand{\company}{JetBrains\xspace}

\AtBeginDocument{%
  }

\copyrightyear{2026}
\acmYear{2026}
\setcopyright{cc}
\setcctype{by}
\acmConference[ICSE '26]{2026 IEEE/ACM 48th International Conference on Software Engineering}{April 12--18, 2026}{Rio de Janeiro, Brazil}
\acmBooktitle{2026 IEEE/ACM 48th International Conference on Software Engineering (ICSE '26), April 12--18, 2026, Rio de Janeiro, Brazil}
\acmPrice{}
\acmDOI{10.1145/3744916.3787811}
\acmISBN{979-8-4007-2025-3/2026/04}

\begin{document}

\title{Evolving with AI: A Longitudinal Analysis of Developer Logs}

\author{Agnia Sergeyuk}
\authornote{Both authors contributed equally to this research.}
\orcid{0009-0001-1495-9824}
\affiliation{%
  \institution{JetBrains Research}
  \city{Belgrade}
  \country{Serbia}
}
\email{agnia.sergeyuk@jetbrains.com}

\author{Eric Huang}
\authornotemark[1]
\orcid{0009-0009-7743-3465}
\affiliation{%
  \institution{University of California, Irvine}
  \city{Irvine}
  \country{United States}}
\email{jianch2@uci.edu}

\author{Dariia Karaeva}
\orcid{0009-0004-4110-0152}
\affiliation{%
 \institution{JetBrains}
 \city{Paphos}
 \country{Cyprus}}
\email{dariia.karaeva@jetbrains.com}

\author{Anastasiia Serova}
\orcid{0009-0002-0914-2294}
\affiliation{%
  \institution{JetBrains}
  \city{Amsterdam}
  \country{Netherlands}}
\email{anastasiia.serova@jetbrains.com}

\author{Yaroslav Golubev}
\orcid{0000-0001-7009-635X}
\affiliation{%
  \institution{JetBrains Research}
  \city{Belgrade}
  \country{Serbia}
}
\email{yaroslav.golubev@jetbrains.com}

\author{Iftekhar Ahmed}
\orcid{0000-0001-8221-5352}
\affiliation{%
  \institution{University of California, Irvine}
  \city{Irvine}
  \country{United States}}
\email{iftekha@uci.edu}

\newcounter{observation}
\newcommand{\observation}[1]{\refstepcounter{observation}
	\begin{center}
		\framebox{
			\begin{minipage}{0.93\columnwidth}
				{} \textit{#1}
			\end{minipage}
		}
	\end{center}
}

\begin{abstract}
AI‑powered coding assistants are rapidly becoming fixtures in professional IDEs, yet their sustained influence on everyday development remains poorly understood. Prior research has focused on short-term use or self-reported perceptions, leaving open questions about how sustained AI use reshapes actual daily coding practices in the long term. 

We address this gap with a mixed-method study of AI adoption in IDEs, combining longitudinal two-year fine-grained telemetry from 800 developers with a survey of 62 professionals. We analyze five dimensions of workflow change: productivity, code quality, code editing, code reuse, and context switching. Telemetry reveals that AI users produce substantially more code but also delete significantly more. Meanwhile, survey respondents report productivity gains and perceive minimal changes in other dimensions.

Our results offer empirical insights into the silent restructuring of software workflows and provide implications for designing future AI-augmented tooling.
\end{abstract}




\maketitle

\newcommand{\llmAgents}{{LLM-based agents }}

\section{Introduction}
\label{section:intro}
Recent advances in Large Language Models (LLMs) have introduced a profound shift in how software is developed~\cite{mozannar2024reading, dell2025cybernetic, tabarsi2025llms}. AI-powered coding assistants such as GitHub Copilot, JetBrains AI Assistant, and Visual Studio IntelliCode are now embedded in the daily workflows of many professional developers~\cite{StackOverflow2024Survey, JetBrains2024DevEcosystem}. These tools are designed to automate routine coding tasks, generate non-trivial code suggestions, and support activities like documentation and test generation~\cite{sergeyuk2025using, liang2024large}. However, while these tools promise productivity and efficiency gains, their long-term effects on how developers work --- how they write, edit, reuse, and navigate code --- remain poorly understood.

Understanding these shifts is not only timely but essential. As AI tools are rapidly adopted, they have the potential to fundamentally restructure developer workflows, with implications for productivity, software quality, team dynamics, and future tooling. Most existing studies on developer-AI interaction rely on short-term experiments~\cite{peng2023impact, imai2022github, vaithilingam2022expectation}, small-scale observations~\cite{dell2025cybernetic, dakhel2023github}, or surveys of developer perceptions~\cite{liang2024large, vaillant2024developers}. While valuable, these approaches provide a limited view of the evolving developer experience. Lab studies may not generalize to the complexity and scale of real-world software development, and self-reported data can differ notably from actual behavior ~\cite{devanbu2016belief, beckermeasuring}. Importantly, most prior work offers only a static view, missing the longer-term evolution of developer practices as they adapt to and co-evolve with AI tools.

We address this gap through a comprehensive, mixed-methods analysis of developer behavior, conducted in collaboration with JetBrains,\footnote{JetBrains: \url{https://www.jetbrains.com/}} a major software vendor specializing in the creation of intelligent development tools. Our study combines longitudinal, large-scale telemetry data with qualitative insights. We analyze 151,904,543 fine-grained, anonymized IDE interaction logs collected over two years from 800 software developers, including 400 \textit{AI users} and 400 \textit{AI non-users}. To complement this behavioral data with self-reported experiences and perspectives, we conducted a survey with 62 professional developers and follow-up interviews with five participants. This approach examines developer workflows through two complementary lenses: what developers do (as observed in logs), and what they perceive (as captured through surveys and interviews), offering a detailed view of AI's real-world impact on software development practices.

The research questions guiding this study are as follows:

\vspace{0.1cm}
\textit{How did the introduction of AI tools influence the...}
\vspace{-0.15cm}
\begin{enumerate}[font={\bfseries},label={RQ\arabic*:},leftmargin=1.25cm]
    \item \textit{...\textbf{productivity} of developers?}
    \item \textit{...\textbf{code quality}?}
    \item \textit{...practices of \textbf{code editing}?}
    \item \textit{...practices of \textbf{code reuse}?}
    \item \textit{...practices of \textbf{context switching}?}
\end{enumerate}

Together, these five dimensions provide a structured way to examine how AI assistance influences not only the outcomes of software development but also the underlying practices and processes. By addressing these five questions, we aim to develop both a fine-grained and holistic understanding of how developer workflows are evolving in response to sustained AI tool usage. 

Our results offer implications for the future of software engineering practice and tooling. 

\begin{itemize}

\item \textbf{Empirical characterization of evolving AI-assisted workflows.} Through examination of logs, we reveal that developers using AI tools write and modify significantly more code, rely more frequently on external sources (e.g., code copied from the web), and switch contexts more often. At the same time, survey respondents perceive minimal changes outside of productivity.

\item \textbf{Reframing of AI's impact on effort and attention.} Our findings show that AI does not uniformly reduce developer effort; instead, it redistributes effort across more fragmented, reactive, and cognitively demanding workflows.

\end{itemize}

Together, these findings deepen our understanding of how AI tools reshape software development in practice. They highlight the limitations of self-reported data, reveal structural changes in workflow that are not always visible to users themselves, and point to new design priorities for AI-assisted environments. Rather than focusing solely on task automation, future tools should also support continuity, reduce cognitive overhead, and enhance developer awareness across fragmented tasks.

The remainder of the paper is organized as follows. Section~\ref{section:related work} reviews relevant literature and motivates the five workflow dimensions studied. Section~\ref{section:methods} describes our mixed-methods approach, including data collection and analysis procedures. Section~\ref{section:result} presents our findings by research questions. Section~\ref{section:discussion} synthesizes the results and their implications. Section~\ref{section:threats} discusses threats to validity, and Section~\ref{section:conclusion} concludes the paper. The supplementary materials for the paper are available online~\cite{artifacts}.

\section{Related Work}
\label{section:related work}
\subsection{Developers' Workflows in IDEs}

Understanding how developers spend their time in IDEs is a focus of interest for both software engineering research and industry. Prior work has leveraged telemetry to uncover behavioral patterns that reflect key aspects of developers' workflow. For instance, Minelli et al. conducted an in-depth analysis based on data mined from developers' interactions within the PHARO IDE~\cite{minelli2015know}. From this, they discovered that developers spend the majority of their time (70\%) on program comprehension activities, which include reading, navigating, and reviewing source code. Similarly, Amann et al.~\cite{amann2016study} analyzed usage traces from Visual Studio, showing that developers interact with the IDE for only 25\% of their time, with frequent brief interruptions punctuating longer work sessions.

To move beyond low-level metrics, researchers have employed sequence modeling and clustering techniques to abstract developers' behaviors. Damevski et al.~\cite{damevski2016interactive} used Hidden Markov Models to categorize actions related to debugging into latent task states. Astromskis et al.~\cite{astromskis2017patterns} analyzed over 1,000 hours of IDE interaction to define six high-level categories --- code-related activities, executing the system under development, using utilities, using external programs, communicating with other developers, and looking for online help. Collectively, these studies demonstrate how telemetry enables scalable, in-situ observation of developer behavior. 

While prior work has provided foundational insights into how developers engage with IDEs, it primarily captures behavior in pre-AI contexts. These studies offer valuable baselines for understanding time allocation, navigation patterns, and debugging workflows, but they do not account for the widespread adoption of AI-powered coding assistants in modern development environments. As such, they leave open critical questions about how AI alters not just developer productivity, but also the structure, cadence, and cognitive demands of everyday development work. Our study addresses this gap by leveraging fine-grained IDE telemetry logs collected over a two-year period from professional developers actively using an AI assistant. Unlike earlier studies focused on static or low-level patterns, we combine these logs with developer survey responses to investigate how behaviors and perceptions evolve over time.

\subsection{AI and Developers' Workflows in IDEs}

The recent rise of LLMs has introduced new and potentially transformative dynamics into software development workflows. AI-based coding assistants, such as GitHub Copilot~\cite{github_copilot} and JetBrains AI Assistant~\cite{jetbrains_ai} offer promises of increased efficiency and reduced manual effort. However, their actual impact on day-to-day development behavior remains unclear. Notably, most prior work has relied on short-term controlled experiments~\cite{peng2023impact, imai2022github, vaithilingam2022expectation} or self-reported perceptions~\cite{liang2024large, vaillant2024developers}, with limited use of longitudinal, telemetry-based evidence. To systematically investigate how AI affects real-world development practices, we focus on five workflow dimensions that are both grounded in prior research and observable through fine-grained telemetry from an enterprise-scale IDE: \textbf{productivity}, \textbf{code quality}, \textbf{code editing}, \textbf{code reuse}, and \textbf{context switching}.

\paragraph{\textbf{Productivity}}
Since the emergence of LLM-based coding tools, numerous studies have explored their impact on developer productivity~\cite{beckermeasuring,imai2022github, vaithilingam2022expectation, weisz2022better, ziegler2022productivity, ross2023programmer, dakhel2023github, coutinho2024role, vaillant2024developers}. However, findings have been mixed, and there remains no clear consensus on whether such tools consistently yield measurable productivity gains in real-world development contexts. Some studies report promising results. For example, \citet{imai2022github} found that pair programming with GitHub Copilot led to increased code output --- as measured by lines added --- suggesting that AI assistance can accelerate the act of writing and modifying code. Similarly, \citet{peng2023impact} demonstrated that Copilot significantly reduced task completion time, yielding over a 50\% improvement in productivity. However, other work urges caution. \citet{vaithilingam2022expectation} observed that Copilot did not consistently reduce task time or improve task success in realistic programming tasks. \citet{ziegler2022productivity} echoed this concern, highlighting a common disconnect between perceived productivity gains and actual improvements in development outcomes. In addition, \citet{beckermeasuring} found that, contrary to developers' own estimates, the use of AI assistants actually increases task completion time by 19\%.

\paragraph{\textbf{Code Quality and Code Editing}}

Prior research has examined how AI assistants are reshaping the programming process, revealing meaningful shifts in developer workflows. Given the imperfect nature of AI-generated code, developers now devote substantial effort to verifying, refining, and integrating AI suggestions. \citet{mozannar2024reading} report that developers spend over 50\% of their time evaluating and editing AI-generated output. Complementary findings from Sahoo et al.~\cite{sahoo2024ansible} show that of the AI-suggested code that is initially accepted, 18.16\% is later deleted and 6.62\% is heavily rewritten --- highlighting the costs of validation and rework.

Developers' attitudes toward AI assistance also vary. Some actively avoid using such tools due to concerns that generated code may not meet functional or non-functional requirements~\cite{liang2024large}. In contrast, others exhibit over-reliance, placing undue trust in AI output and showing reduced willingness to edit or critically assess it~\cite{prather2023s, schaffer2019can}. This variation raises concerns about both under- and over-utilization of AI in the development process.

The impact of AI-generated code on code quality remains similarly inconclusive. Some studies document performance-related inefficiencies introduced by LLMs, such as deeply nested loops and redundant variable declarations~\cite{lertbanjongngam2022empirical}. However, other findings suggest that these issues may not substantially degrade the overall code quality. For example, Song et al.~\cite{song2024impact} report no significant change in quality despite observable productivity improvements. \citet{weisz2022better} further argue that even when AI suggestions are suboptimal, they can function as effective scaffolding, shifting the task from code generation to code review.

\paragraph{\textbf{Code Reuse}}

Code reuse --- the practice of leveraging existing code, libraries, APIs, or prior implementations rather than writing functionality from scratch --- has long been a cornerstone of efficient software development. However, the advent of AI-powered coding tools is reshaping reuse practices in subtle but significant ways. These tools frequently assist developers by generating boilerplate code and suggesting commonly used patterns or snippets derived from training data~\cite{Chen2021}. Recent empirical studies suggest that AI tools may both enable and displace traditional reuse strategies. \citet{liang2024large} found that developers often rely on AI assistants for generating repetitive structures and simple logic, potentially reducing their dependence on existing libraries or shared components. At the ecosystem level, \citet{del2023large} reported a notable drop in StackOverflow activity --- from approximately 60,000 to 40,000 new posts per month --- within six months of ChatGPT release, suggesting a shift away from traditional forms of external knowledge reuse. Moreover, AI-generated code may unintentionally ``reinvent the wheel'' by replicating functionality already present in the codebase or available through standard libraries~\cite{sergeyuk2024design}. This raises concerns about the efficiency and sustainability of AI-facilitated reuse, particularly when such tools obscure opportunities to leverage mature, well-tested components.

\paragraph{\textbf{Context Switching}}

One of the key promises of integrated AI coding assistants is their potential to reduce context switching by minimizing the need to leave the IDE to search for external solutions. By surfacing relevant suggestions and code completions inline, these tools aim to make development workflows more seamless and less interruptive~\cite{pandey2024transforming, sergeyuk2024design}. However, emerging evidence suggests that new forms of disruption may offset these benefits. For example, \citet{weber2024significant} report that interacting with AI interfaces can introduce additional cognitive load and increase task fragmentation, as developers must frequently shift between writing code, interpreting suggestions, and managing communication with the AI system. These findings raise questions about whether AI assistance genuinely reduces context switching or replaces one form of interruption with another.

\paragraph{\bf Summary}

Despite the growing number of studies on AI in software engineering, the overall picture remains inconclusive. While some report perceived gains in productivity or faster code authoring, others highlight increased verification overhead or decreased code quality, resulting in a decrease in productivity. These mixed findings often stem from differences in study design, developer expertise, and task complexity, resulting in a fragmented and sometimes contradictory understanding of how AI tools reshape real-world development workflows.

To address this gap, we present a study of how AI tools affect developer workflows over a two-year period. Using anonymized IDE telemetry data from professional developers, combined with a survey capturing developer perceptions, we offer a comprehensive view of this transformation. Our analysis is grounded in five theoretically informed and operationally measurable dimensions of workflow. By complementing objective behavioral data with self-reported experiences, our study provides an integrated and empirically grounded perspective on how AI tools are reshaping software development practice over time.

\section{Methodology}
\label{section:methods}
To study the influence of AI on developer workflows from different angles, we formulated research questions that focus on \textbf{how AI tools influence \textit{productivity}, \textit{code quality}, as well as the practices of \textit{code editing}, \textit{code reuse}, and \textit{context switching}}. We performed a mixed-method study, consisting of two stages:
\begin{itemize}
    \item \textbf{survey} of developers and the \textbf{post-survey interviews} that provided contextual insights;
    \item \textbf{log analysis} of actual developer behavior in the IDE.
\end{itemize}

This mixed-method approach examines AI adoption through two complementary but distinct lenses. Surveys capture retrospective, holistic evaluations of workflow change. Logs capture moment-to-moment interactions at the event level, aggregated monthly across thousands of actions. These sources measure different constructs and address the research questions from different temporal and granular perspectives. Our goal is to understand what developers actually do in their IDEs and what they perceive about those activities. Below, we describe each of the stages in detail.

\subsection{Survey and Post-Survey Interviews}
\label{sec:methodology:survey}

Firstly, to gather the self-reported perceptions from practitioners, we carried out an online survey aimed at our stated research questions. The study was conducted in line with our institution's ethical standards, adhering to the values and guidelines outlined in the ICC/ESOMAR International Code~\cite{iccesomar}.

\subsubsection{Data collection}

\textbf{Questions.} 
Our survey consists of:
\begin{itemize}
    \item Five \textbf{demographic questions}.
    \item Two 5-point Likert-scale questions about the \textbf{overall experience} with AI tools for coding and the \textbf{reliance} on them.
    \item Seven 5-point Likert-scale questions regarding \textbf{perception of their workflow} after they started using AI. Specifically, the questions focus on the changes to the five aspects of our study (from \textit{Significantly decreased} to \textit{Significantly increased}):

    \begin{itemize}
        \item For \textbf{productivity}, we asked directly about \textbf{overall productivity} and also about \textbf{time spent coding} (\textbf{RQ1)}.
        \item For \textbf{code quality}, we asked directly about the \textbf{quality of code} and also about \textbf{code readability} (\textbf{RQ2)}.
        \item For \textbf{code editing}, we asked about the \textbf{frequency of editing or modifying own code} (\textbf{RQ3)}.
        \item For \textbf{code reuse}, we asked about the \textbf{frequency of use of code from external sources}, \textit{e.g.}, libraries, online examples, or AI-suggested code (\textbf{RQ4)}.
        \item For \textbf{context switching}, we asked directly about the \textbf{frequency of context switching} --- switching between different tasks or thought processes (\textbf{RQ5)}.
    \end{itemize}
    
    \item An open-ended question where respondents were allowed to write a \textbf{specific example} of how AI tools for coding have impacted their workflow. 
    \item A question asking respondents if they want to participate in a short \textbf{post-survey interview}.
\end{itemize}

To ensure question clarity and unambiguity, we conducted two rounds of expert review with the institutional copyediting team and five test runs with colleague developers from JetBrains to validate survey question comprehension.
The full questionnaire can be found in supplementary materials~\cite{artifacts}.

\textbf{Participants.} Recruitment was conducted via a survey link sent to a subset of developers who had previously consented to be contacted for research purposes as part of an internal participant panel maintained by \company. This subset was selected based on prior survey responses in which they self-identified as users of AI-assisted development tools. These prior responses were collected as part of separate user studies conducted by \company, with explicit consent for data retention and recontact. In line with internal research governance policies, no participants were contacted more than once within a 30-day period. A total of 1,231 invitation emails were sent to eligible participants, yielding 76 link clicks and ultimately 67 completed survey responses. As a thank you, participants were given the opportunity to enter a draw for USD 50 Amazon eGift Cards or an equivalent-value \company product pack. 

Overall, we received 67 complete survey responses. After excluding participants who indicated that they had not used AI assistance for coding, 62 responses remained for analysis. Of them, 56 identified as \textit{Developers}, 16 as \textit{Team Leads}, 10 as \textit{Architects}, and 10 as \textit{DevOps Engineers}. Participants could select up to three roles that described them. In terms of professional coding experience, 22 participants had \textit{16 or more years}, 14 had \textit{11–15 years}, 17 had \textit{6–10 years}, 8 had \textit{3–5 years}, and 1 had \textit{1–2 years}.

Regarding AI assistance tools, the three most commonly used were \textit{ChatGPT} (\(n=42\)), \textit{GitHub Copilot} (\(n=29\)), and \textit{JetBrains AI Assistant} (\(n=23\)). Participants could select multiple tools. The distribution of AI tool usage duration was as follows: 28 participants had used AI assistance for \textit{more than one year}, 12 for \textit{7–12 months}, 12 for \textit{4–6 months}, 6 for \textit{2–3 months}, and 4 for \textit{less than one month (1–4 weeks)}.

\textbf{Post-survey interviews.} In our survey, we asked participants for their consent to contact them for a post-survey interview to get more context on their answers. We selected five interview participants from survey respondents (P7, P15, P33, P39, and P62) who indicated interest in follow-up interviews. We aimed to capture diverse perspectives to ensure representative insights. We diversified across experience levels (3-5 years to 16+ years), geographic locations (Serbia, Chile, UK, Czech Republic, Argentina), AI tool satisfaction levels (from Very Dissatisfied to Very Satisfied), and usage patterns (regular users, occasional users, and those who stopped using AI tools). The five selected participants also represented different professional roles (Developer, DevOps Engineer, Architect, CIO), and AI tool usage durations (from weeks to over a year). This sampling approach captured varying experiences with AI adoption, including both positive and negative perspectives. 

These participants were additionally rewarded with USD 50 Amazon eGift Cards or an equivalent-value \company product pack upon their participation in the short (30 min), semi-structured interviews. The interviews focused on why participants reported specific perceptions and their reasoning behind them. We had \company's copyeditors review the interview script and performed a pilot session to refine question formulations and ensure clarity of the semi-structured interview approach. All interviews were conducted in English via Google Meet.\footnote{Google Meet: \url{https://meet.google.com/landing}} The full interview script can be found in supplementary materials~\cite{artifacts}.

We selected interview participants from survey respondents who indicated interest in follow-up interviews. We aimed to capture diverse perspectives to ensure representative insights. We diversified across experience levels (3-5 years to 16+ years), geographic locations (Serbia, Chile, UK, Czech Republic, Argentina), AI tool satisfaction levels (from Very Dissatisfied to Very Satisfied), and usage patterns (regular users, occasional users, and those who stopped using AI tools). The five selected participants represented different professional roles (Developer, DevOps Engineer, Architect, CIO), and AI tool usage durations (from weeks to over a year). This sampling approach captured varying experiences with AI adoption, including both positive and negative perspectives. 

\subsubsection{Data analysis}

To characterize users' perceptions of changes in their workflows, we calculated descriptive statistics for the questions regarding productivity, code quality, code editing, code reuse, and context switching. 

For the answers to the open-ended question, we familiarized ourselves with their content and, based on it, invited participants for further short interviews. The interviews were transcribed into text using Google Meet's built-in function and then validated by the authors. Similar to the answers to open-ended questions in the survey, the authors of the paper familiarized themselves with the transcripts and selected quotes that provided deeper context for the answers to the survey questions and overall findings.

\subsection{Log Analysis}

To examine how developer behavior evolves over time, we carried out an analysis of IDE usage logs.

\subsubsection{Data collection}

The data about developer activities in the IDEs was provided by \company in an anonymized form and consisted of the logs of specific actions (see details below). Each log consisted of a unique device ID (corresponding to a user), the name of the event, its timestamp, and various metadata. 

\textbf{Sample}. The data consisted of usage logs from several IDEs, namely IntelliJ IDEA, PyCharm, PhpStorm, and WebStorm. Importantly, the studied IDEs covered different programming languages and ecosystems, representing a varied population of IDE users, not beholden to one particular language.

We gathered our samples from devices, for which the following two conditions were true:
\begin{itemize}
    \item The device had some IDE activity in October of 2022.
    \item The device had some IDE activity in October of 2024.
\end{itemize}
The starting point of October 2022 was chosen because the first version of ChatGPT was released in November of 2022~\cite{openai2022chatgpt}, an event that enabled researchers and tool builders to start integrating LLMs into developers' tooling. For all the devices, we have the data for all the two years between these dates, allowing us to carry out a longitudinal analysis of changes that occurred over this time month-by-month. 

We further draw two equal samples from this data:

\begin{itemize}
    \item \textbf{AI users}: 400 devices that consistently interacted with JetBrains AI Assistant\footnote{JetBrains AI Assistant: \url{https://www.jetbrains.com/ai-assistant/}. This product is a general-purpose in-IDE AI assistant that integrates code completion, AI chat, code generation, and documentation features, capturing a diverse range of interaction patterns with AI inside IDEs.} every month, at least once per month, from April 2024 to October 2024. 
    \item \textbf{AI non-users}: 400 devices that never interacted with JetBrains AI Assistant during the studied two years.
\end{itemize}
The condition of using the in-IDE JetBrains AI Assistant since at least April 2024 allows us to ensure that \textit{AI users} are people who have eventually adopted AI assistance into their development workflow. This window coincides with the period when AI assistants became widely available and stable in professional development environments, ensuring our \textit{AI users} sample represents authentic adopters, who have integrated AI into their regular workflow rather than occasional experimenters. This allows us not only to look at logs but also compare AI adopters with other users, seeing how the rate of change for the studied aspects differed between the groups. 

\textbf{Metrics and actions}. IDE telemetry logs are inherently complex and noisy, making it challenging to extract reliable insights without careful abstraction. To obtain robust and interpretable results, we focus on general, well-defined user actions that are consistently captured across sessions. Accordingly, we operationalize each of the five workflow aspects --- productivity, code quality, code editing, code reuse, and context switching --- in terms of measurable in-IDE behaviors recorded via telemetry.

While the generality of these actions means we cannot directly attribute observed changes solely to AI features, it is reasonable to assume that the introduction and increasing use of AI assistants would influence these behaviors. Our goal is not to isolate causal effects, but rather to detect patterns of change in developer workflows over time, as AI tools become more integrated into everyday development practice.
Moreover, having two samples of \textit{AI users} and \textit{AI non-users} allows us to see whether they statistically significantly differed in these general metrics. 

Here are the actions and metrics we collected for each aspect:

\begin{itemize}
    \item As a proxy for \textbf{productivity}, we counted the \textbf{number of typed characters in the editor (RQ1)}, reflecting the granularity of our available telemetry data. While code output has been used as an indicator of developer activity~\cite{forsgren2021space, murphy2019predicts}, we acknowledge that it captures only one specific facet of productivity --- namely, code authoring --- and does not account for other important activities such as debugging, design, or collaboration. Nonetheless, it serves as a consistent and interpretable measure for analyzing changes in active code-writing behavior over time.
    \item As a proxy for \textbf{code quality}, we chose the \textbf{number of debugging instances (RQ2)}. Specifically, we collected all cases when developers started debugging (either by a hotkey or through contextual IDE actions). An increase in debugging activity may signal a rise in code defects or uncertainty, suggesting that developers are encountering more issues that require investigation. 
    \item As a proxy for \textbf{code editing}, we chose the \textbf{number of deletions}, including delete keystrokes, backspaces, undos, etc. \textbf{(RQ3)}. This aggregate measure reflects the frequency with which developers modify or iterate on existing code, providing insight into the intensity of editing activity. While it does not distinguish between types of edits, it serves as a general indicator of the effort devoted to refining code. 
    \item As a proxy for \textbf{code reuse}, we chose the \textbf{instances of external content pasted into the IDE (RQ4)}. While paste actions can have diverse motivations and sources, the recent advancements in AI assistants, particularly chat-based interfaces, have increased copying from external suggestions~\cite{harding2025aicodequality}. For this RQ, we measure external code integration behavior rather than source-specific attribution. Therefore, we focused on paste actions that were not preceded by a copy action. This suggests that the content was not copied from within the codebase in the IDE but originated from external sources. This operationalization allows us to approximate the reuse of external code or AI-generated snippets. 
    \item As a proxy for \textbf{context switching}, we chose the \textbf{number of IDE window activations (RQ5)}. This action refers to cases when the IDE was inactive (meaning the user was focused on another window, \textit{e.g.}, a browser), and then they switched back to the IDE. This metric captures how frequently developers shift their attention between the programming environment and external sources of information, providing insight into workflow fragmentation and interruptions. 
\end{itemize}

Analyzing the dynamics of these actions enables a systematic observation of workflow evolution in \textit{AI users} and \textit{AI non-users}. Overall, our dataset comprised 151,904,543 logged events performed by the studied 800 users.

\subsubsection{Data processing}

We collected longitudinal IDE telemetry data from October 2022 to October 2024 across all devices and conducted an analysis to understand how user behavior evolves over time. The raw data consisted of timestamped records of individual user actions. To reveal broader trends from this granular data, we aggregated actions on a monthly basis for each device.

Specifically, for each device and month, we computed the total count of occurrences for each tracked action.  For example, we could determine that: \textit{User A performed `Action X' 15 times in January, 2023, and 22 times in February, 2023.} This process resulted in a clear, high-level dataset of monthly counts for every action, which is ideal for tracking activity and identifying meaningful behavioral trends.

No extensive filtering was applied to the data. In the cases when a particular user did not perform a particular action in the given month, we added it with a value of 0 to the list described above. This step ensured consistency across devices and time windows, preserving the full sequence for analysis.

\subsubsection{Data analysis}

Having obtained the final aggregated data, we moved on to analyzing it.

\textbf{Normality and homogeneity checks.} To inform the choice of statistical models, we tested for normality and homogeneity of variances. The Kolmogorov–Smirnov test~\cite{scipy_kstwo} indicated that the distribution of action counts per device is not normal for all metrics. Bartlett's test~\cite{scipy_bartlett} showed that variances were not equal across groups and time points for most metrics. These results helped us choose the correct statistical tests.

\textbf{Trend analysis.} Having the aggregated monthly data for each metric, we carried out trend analysis on the longitudinal range from October 2022 to September 2024 (to cover exactly two years, 24 months). Given the non-normal distribution and heteroscedasticity of our data, we performed Mixed Linear Model Regression~\cite{statsmodels_mixedlm} for trend analysis. This approach is well-suited for data with repeated measures, where multiple observations are nested within subjects (in our case, devices). It allowed us to estimate changes over time, while accounting for inter-device variability. We fit a separate mixed-effects model for each of the five action types using the following formula:

\[count_{action} \sim group \times n_{month} + (1 | id_{device})\]

Here, \textit{group} indicates whether the device belongs to an \textit{AI user} or an \textit{AI non-user}, $n_{month}$ is the numeric index of the month, and $(1 | id_{device})$ specifies a random intercept for each device. This model enabled us to assess:
\begin{itemize}
    \item Baseline differences between groups (\textit{AI users} vs \textit{AI non-users}).
    \item Overall temporal trends (\textit{e.g.}, whether the frequency of an action increases or decreases over time).
    \item Interaction effects, \textit{i.e.}, whether the trends differ significantly between groups.
\end{itemize}

Statistical significance was assessed using p-values, with a threshold of $p < 0.05$. You can find the full outputs of statistical testing, which highlight that our results are statistically significant, in supplementary materials~\cite{artifacts}.

Overall, this analysis enables a comparative examination of \textit{AI users} and \textit{AI non-users} with respect to how their coding practices evolve over time. We examine these telemetry patterns alongside survey responses to provide complementary views of AI adoption expressed in practitioners' perceptions of AI's impact and the actual changes observed in their development workflows.

\section{Results}
\label{section:result}

In this section, we go over each of the five key aspects (\textbf{RQs 1---5}), drawing on the data from all the sources: survey, interviews, and log analysis.

Before focusing on the aspects, we also asked survey participants about their overall satisfaction with AI coding tools and their changing levels of reliance over time. The responses were largely positive. In terms of satisfaction, 71\% of participants reported being satisfied or very satisfied, while only 11.3\% reported dissatisfaction. Similarly, when asked whether their reliance on AI tools had changed since first adopting them, 61.3\% indicated that their reliance had slightly or significantly increased, whereas only 14.5\% reported a decrease. These results suggest that developers generally view AI tools favorably. However, such aggregate perceptions may obscure important nuances in how AI affects specific aspects of development work. To unpack these subtleties, we now examine perceptions around individual coding practices and juxtapose them with objective behavioral data from telemetry logs to answer the RQs. We present the findings in the following subsections.

 \subsection{RQ1: Productivity}

\begin{figure}[t]
  \centering

  \begin{subfigure}{\linewidth}
    \centering
    \includegraphics[width=\linewidth]{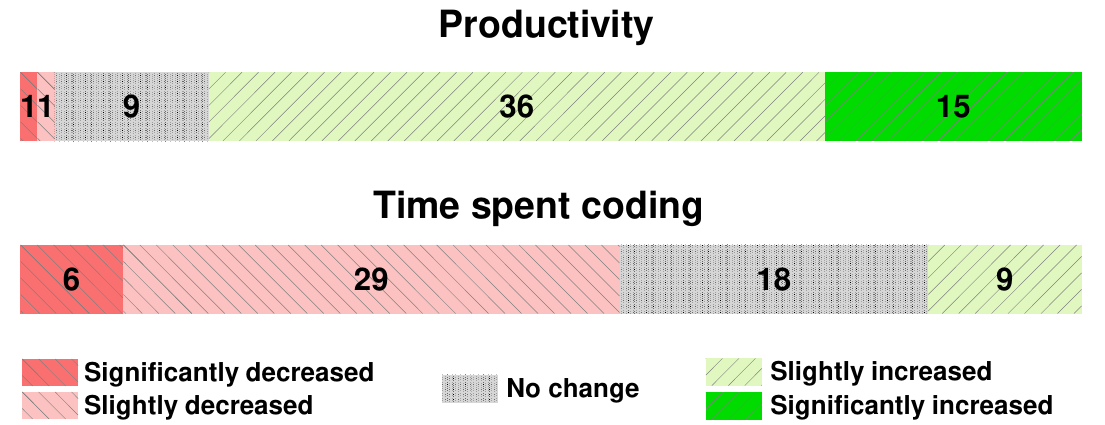}
    \caption{Responses to the survey questions whether \textit{Productivity} and \textit{Time spent coding} decreased or increased due to using AI tools for coding.}
    \label{fig:rq1:survey}
  \end{subfigure}

  \vspace{1em} 

  \begin{subfigure}{\linewidth}
    \centering
    \includegraphics[width=\linewidth]{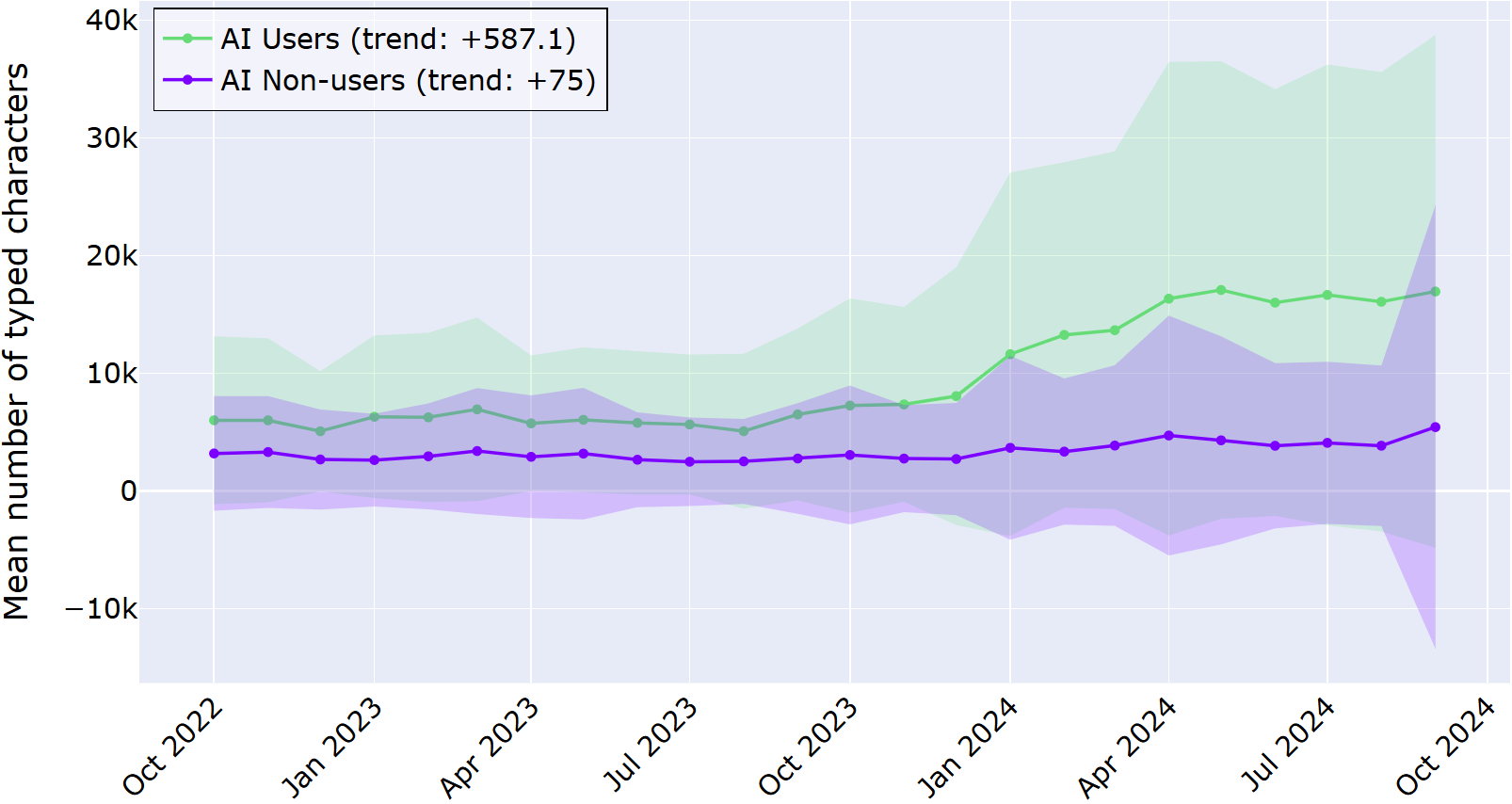}
    \caption{The mean \textit{Number of typed characters} per device in the IDE logs, for different months. \textit{AI Users} show a statistically significant increasing trend of +587 characters a month, \textit{AI Non-users} --- +75 characters. Shaded regions represent ±1 standard deviation from the monthly mean.}
    \label{fig:rq1:logs}
  \end{subfigure}

  \caption{Results for RQ1: Productivity.}
  \label{fig:rq1}
\end{figure}

The results for \textbf{productivity} are shown in Figure~\ref{fig:rq1}. Below, we first present findings from the survey responses, followed by insights derived from the telemetry log analysis.

\textbf{Survey and interviews} (Figure~\ref{fig:rq1:survey}). 
For 82.3\% of respondents, the introduction of AI coding tools slightly or significantly increased their productivity, while two respondents said that it slightly or significantly decreased. Additionally, we asked about the influence of AI coding tools on time spent coding. Here, the results corroborate the same narrative: 56.5\% of respondents said that their coding time decreased, and 14.5\% indicated that it increased. 

Judging these answers together, we can see the overall positive relation of participants towards productivity with AI-based tools. P7, a developer with 3-5 years of experience who regularly uses AI,  provided an example of this:

\begin{quote}
    \textit{``When I get stuck on naming or documentation, I immediately turn to AI, and it really helps.''}
\end{quote}

P33, a developer with more than 10 years of professional experience who churned from AI usage, on the other hand, provides a different testimony:

\begin{quote}
    \textit{``I see a lot of time wasted trying to craft the right prompt to get the desired solution.''}
\end{quote}

\textbf{Logs} (Figure~\ref{fig:rq1:logs}).  
Telemetry data reveal that the number of typed characters used as a proxy for productivity increased over time for both \textit{AI users} and \textit{AI non-users}, with both trends being statistically significant. However, the rate of increase differs distinctly between the two groups. According to our mixed-effects linear model, \textit{AI non-users} showed an average increase of 75 characters per month, whereas \textit{AI users} exhibited a much steeper rise of 587 characters per month. This substantial difference suggests that AI adoption is associated with a greater acceleration in active code authoring over time.

\observation{\textbf{Summary of RQ1}. Respondents report strong perceived productivity gains with the introduction of the AI-based tools. The logs show that \textit{AI users} are writing more code than \textit{AI non-users}.}

\subsection{RQ2: Code Quality}
\begin{figure}[t]
  \centering

  \begin{subfigure}{\linewidth}
    \centering
    \includegraphics[width=\linewidth]{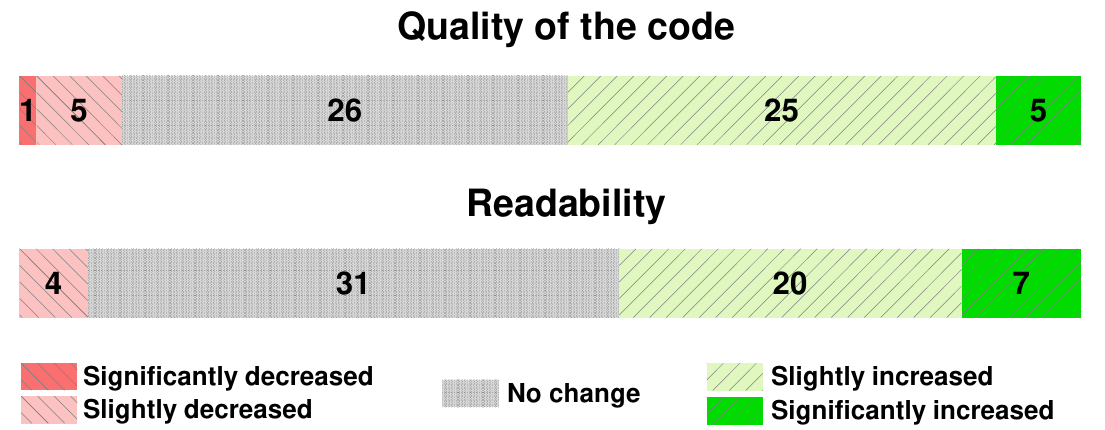}
    \caption{Responses to the survey questions whether \textit{Quality of the code} and \textit{Readability} decreased or increased due to using AI tools for coding.}
    \label{fig:rq2:survey}
  \end{subfigure}

  \vspace{1em} 

  \begin{subfigure}{\linewidth}
    \centering
    \includegraphics[width=\linewidth]{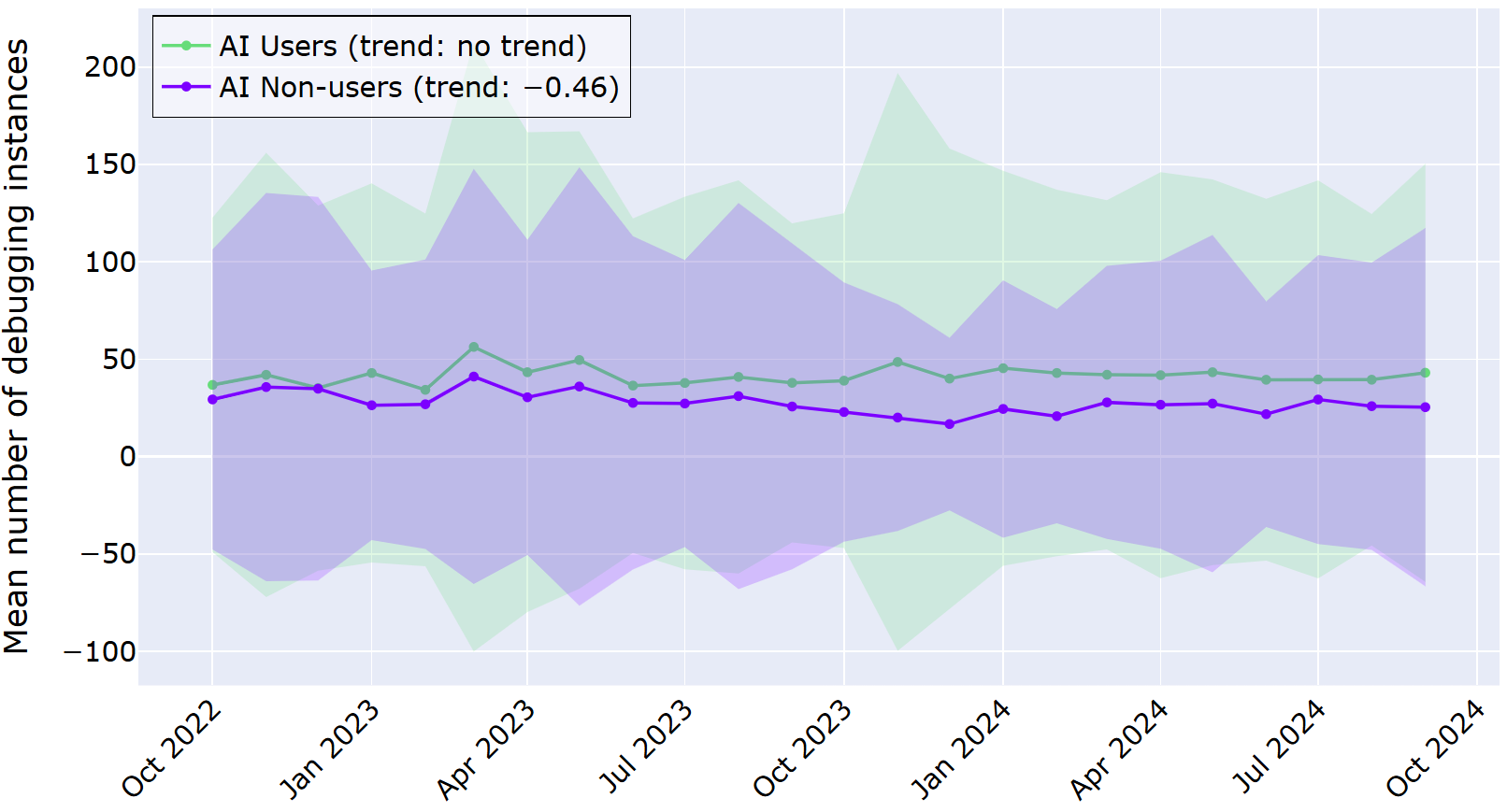}
    \caption{The mean \textit{Number of debugging instances} per device in the IDE logs, for different months. \textit{AI Users} show no trend, \textit{AI Non-users} --- a statistically significant decrease of $-$0.46 instances a month. Shaded regions represent ±1 standard deviation from the monthly mean.}
    \label{fig:rq2:logs}
  \end{subfigure}

  \caption{Results for RQ2: Code quality.}
  \label{fig:rq2}
\end{figure}

The results for \textbf{code quality} are shown in Figure~\ref{fig:rq2}.

\textbf{Survey and interviews} (Figure~\ref{fig:rq2:survey}). When asked if the quality of their code increased because of using AI coding tools, 48.4\% of respondents say that it slightly or significantly increased, while 9.7\% say that it slightly or significantly decreased. For the readability of the code, the respective numbers are 43.5\% and 6.5\%, though 50\% indicate that they did not observe a change. Overall, it can be seen that the view of the change in code quality is somewhat positive. However, P7 highlighted an interesting point about the relationship towards the quality of generated code:

\begin{quote}
    \textit{``If something was done by AI, I'm actually more paranoid about it. I keep wondering, "Is this really correct?" I triple-check it, and even then, I still feel a bit uneasy.''}
\end{quote}

\textbf{Logs} (Figure~\ref{fig:rq2:logs}). We can see that for \textit{AI users} the number of debugging events in logs has no statistically significant trend, meaning no noticeable change in the studied two years. In contrast, \textit{AI non-users} exhibit a statistically significant decline in debugging activity, with an average decrease of approximately 0.46 debugging events per month (\textit{p} < 0.001). This suggests that while debugging behavior remains stable among \textit{AI users}, \textit{AI non-users} are gradually engaging in less debugging over time. 

\observation{\textbf{Summary of RQ2}. Survey respondents report moderately positive perceptions of code quality after using AI-based coding tools, though half perceive no change. Telemetry shows stable debugging activity among \textit{AI users} and declining debugging activity among \textit{AI non-users}.}

\subsection{RQ3: Code Editing}

\begin{figure}[t]
  \centering

  \begin{subfigure}{\linewidth}
    \centering
    \includegraphics[width=\linewidth]{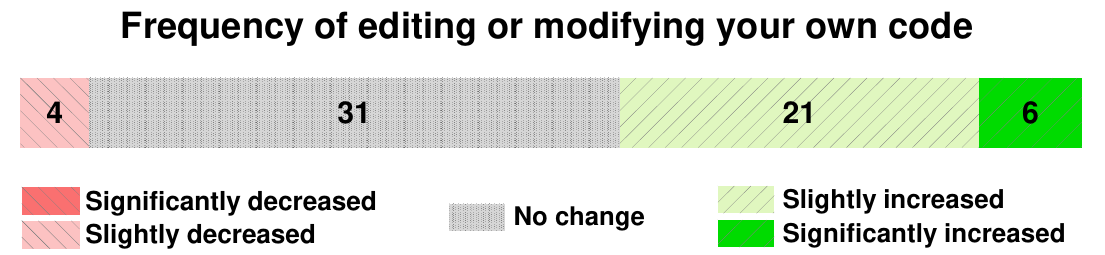}
    \caption{Responses to the survey question whether the \textit{Frequency of editing or modifying your own code} decreased or increased due to using AI tools for coding.}
    \label{fig:rq3:survey}
  \end{subfigure}

  \vspace{1em} 

  \begin{subfigure}{\linewidth}
    \centering
    \includegraphics[width=\linewidth]{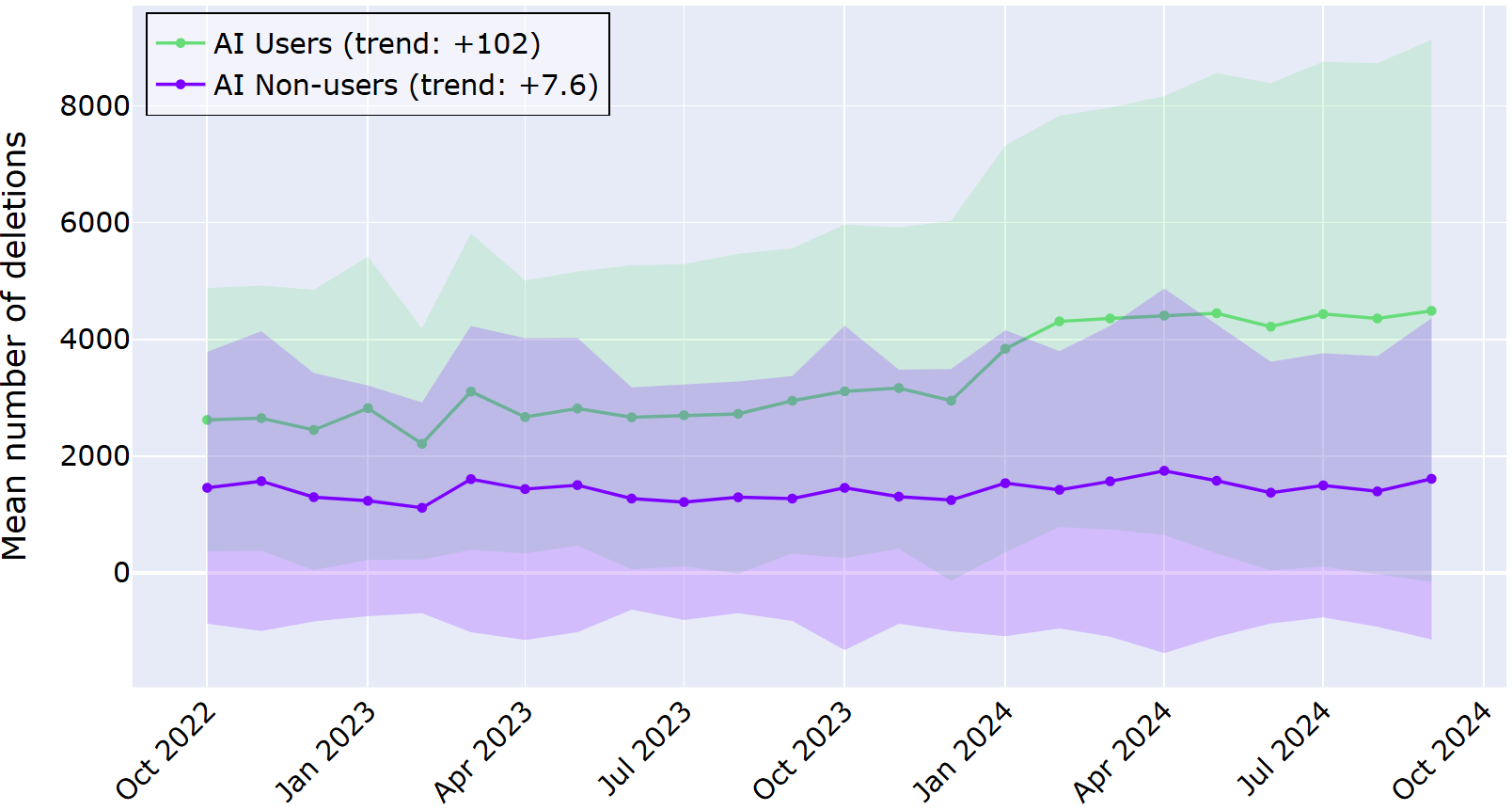}
    \caption{The mean \textit{Number of deletions} per device in the IDE logs, for different months. \textit{AI Users} show a statistically significant increasing trend of +102 deletions a month, \textit{AI Non-users} --- +7.6 deletions. Shaded regions represent ±1 standard deviation from the monthly mean.}
    \label{fig:rq3:logs}
  \end{subfigure}

  \caption{Results for RQ3: Code editing.}
  \label{fig:rq3}
\end{figure}

The results for \textbf{code editing} are shown in Figure~\ref{fig:rq3}. 

\textbf{Survey and interviews} (Figure~\ref{fig:rq3:survey}). Our results show that 43.5\% of participants feel that the frequency of editing their code slightly or significantly increased with the usage of AI tools for coding, 6.5\% feel like it decreased, and 50\% believe it did not change.

P62, a system architect with more than 16 years of experience, noted that the regular use of AI tooling supports their code editing and reviewing practices:
\begin{quote}
    \textit{``AI is like a second pair of eyes, offering pair programming benefits without social pressure --- especially helpful for neurodivergent people. It's not always watching, but I can call on it for code review and feedback when needed.''}
\end{quote}

\textbf{Logs} (Figure~\ref{fig:rq3:logs}). The analysis of deletions in the usage logs shows that, similar to typing, the number of deletions increased for both \textit{AI users} and \textit{AI non-users}, but it increased significantly faster for \textit{AI users}, with an average rate of 102 deletions per month (whereas \textit{AI non-users} showed an increase of about 7.6 deletions per month).

\observation{\textbf{Summary of RQ3}. Survey responses indicate that developers perceive little to no change in their code editing habits with AI, though some report a slight increase. Telemetry data reveals a clear rise in code deletions over time among \textit{AI users}.}

\subsection{RQ4: Code Reuse}

\begin{figure}[t]
  \centering

  \begin{subfigure}{\linewidth}
    \centering
    \includegraphics[width=\linewidth]{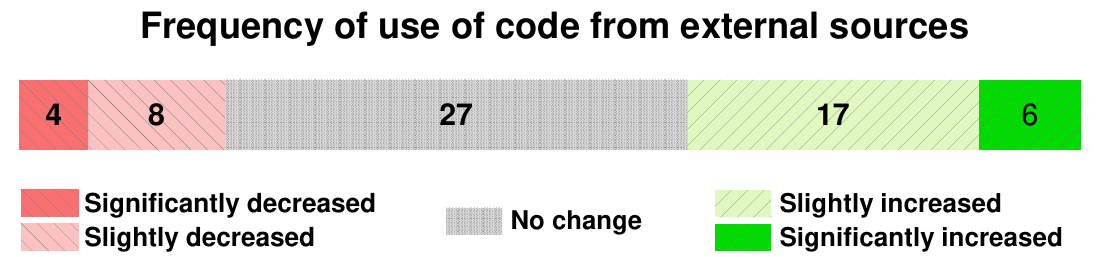}
    \caption{Responses to the survey question whether the \textit{Frequency of use of code from external sources} decreased or increased due to using AI tools for coding.}
    \label{fig:rq4:survey}
  \end{subfigure}

  \vspace{1em} 

  \begin{subfigure}{\linewidth}
    \centering
    \includegraphics[width=\linewidth]{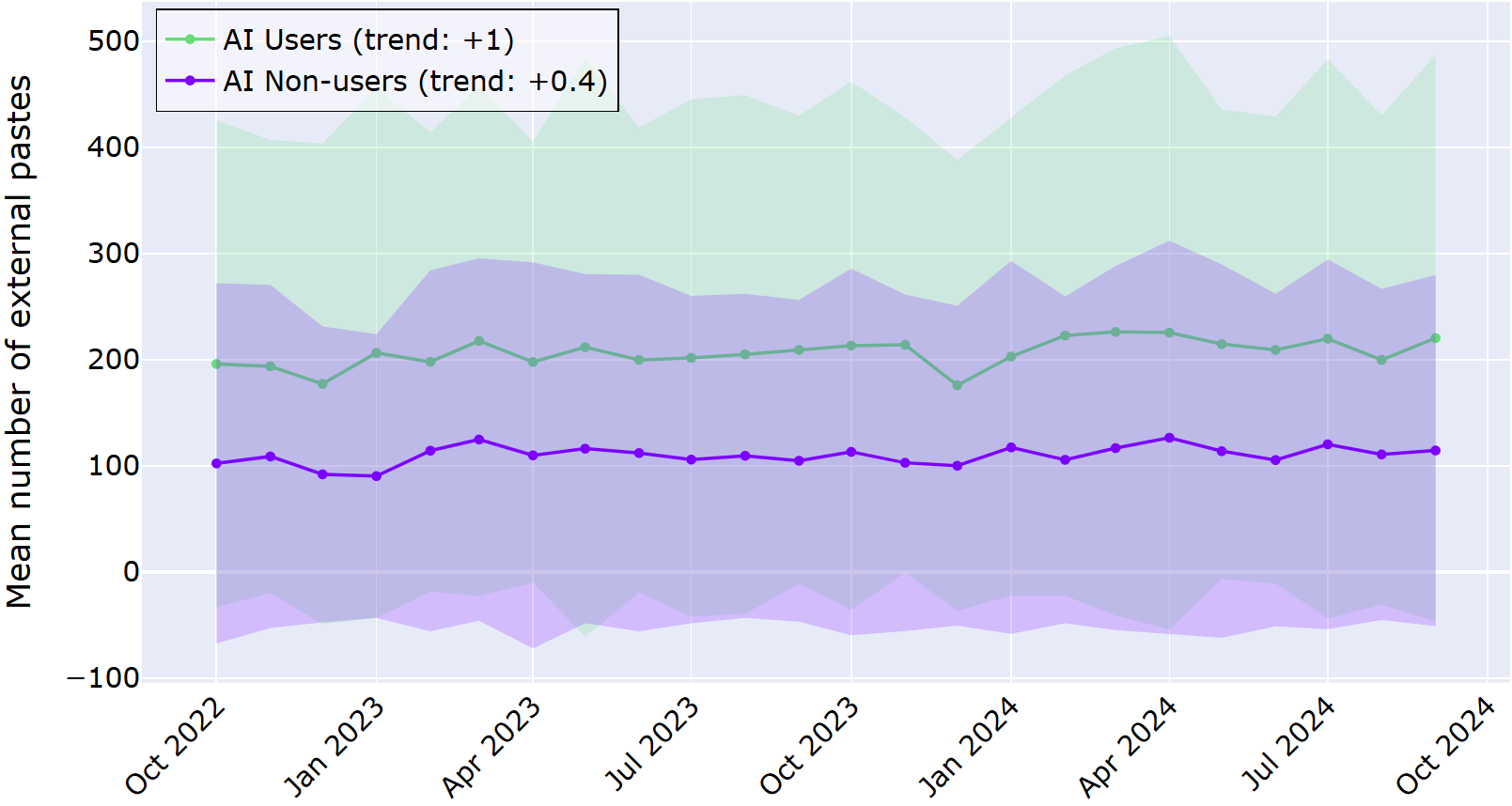}
    \caption{The mean \textit{Number of external pastes} per device in the IDE logs, for different months. \textit{AI Users} show a statistically significant increasing trend of +1 paste a month, \textit{AI Non-users} --- +0.4 pastes. Shaded regions represent ±1 standard deviation from the monthly mean.}
    \label{fig:rq4:logs}
  \end{subfigure}

  \caption{Results for RQ4: Code reuse.}
  \label{fig:rq4}
\end{figure}

The results for \textbf{code reuse} are presented in Figure~\ref{fig:rq4}. 

\textbf{Survey and interviews} (Figure~\ref{fig:rq4:survey}). 37.1\% of respondents reported that their use of code from external sources slightly or significantly increased with the adoption of AI tools, while 19.4\% said it decreased, and 43.5\% observed no change. This distribution is notable, as code reuse might be expected to follow a more consistent trend. Instead, developers report divergent experiences. 
Representing the ``decreasing'' perspective, P39, a DevOps engineer with over 16 years of experience who discontinued AI tool usage, explained:

\begin{quote}
    \textit{``For me, it's better to take responsibility for what I did myself rather than adopt a third-party solution, try to understand it, and convince myself it's correct.''}
\end{quote}

\textbf{Logs} (Figure~\ref{fig:rq4:logs}). In terms of actions of pasting from other sources in logs, we found a small, but statistically significant increase over time for both \textit{AI users} and \textit{AI non-users}. However, it is faster for \textit{AI users} (\textit{p} = 0.03). \textit{AI users} show on average an increase of 1 paste per month, while \textit{AI non-users} --- 0.4 pastes per month.

\observation{\textbf{Summary of RQ4}.
For code reuse, participant opinions on changes in frequency were relatively balanced, with no clear consensus. Telemetry reveals that paste actions from external sources increased over time for both groups, more prominently among \textit{AI users}.
}

\subsection{RQ5: Context Switching}

\begin{figure}[t]
  \centering

  \begin{subfigure}{\linewidth}
    \centering
    \includegraphics[width=\linewidth]{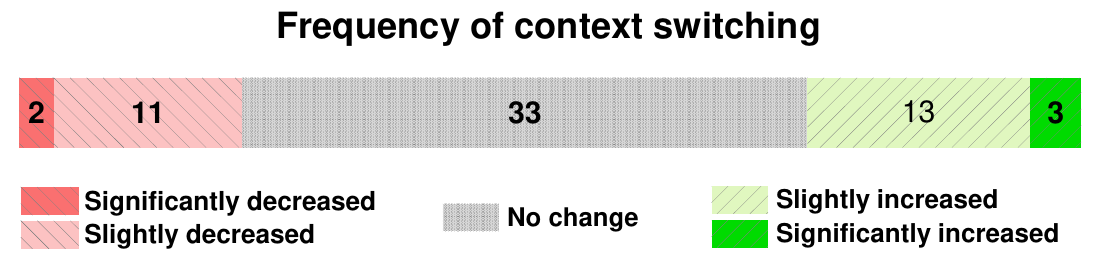}
    \caption{Responses to the survey question whether the \textit{Frequency of context switching} decreased or increased due to using AI tools for coding.}
    \label{fig:rq5:survey}
  \end{subfigure}

  \vspace{1em} 

  \begin{subfigure}{\linewidth}
    \centering
    \includegraphics[width=\linewidth]{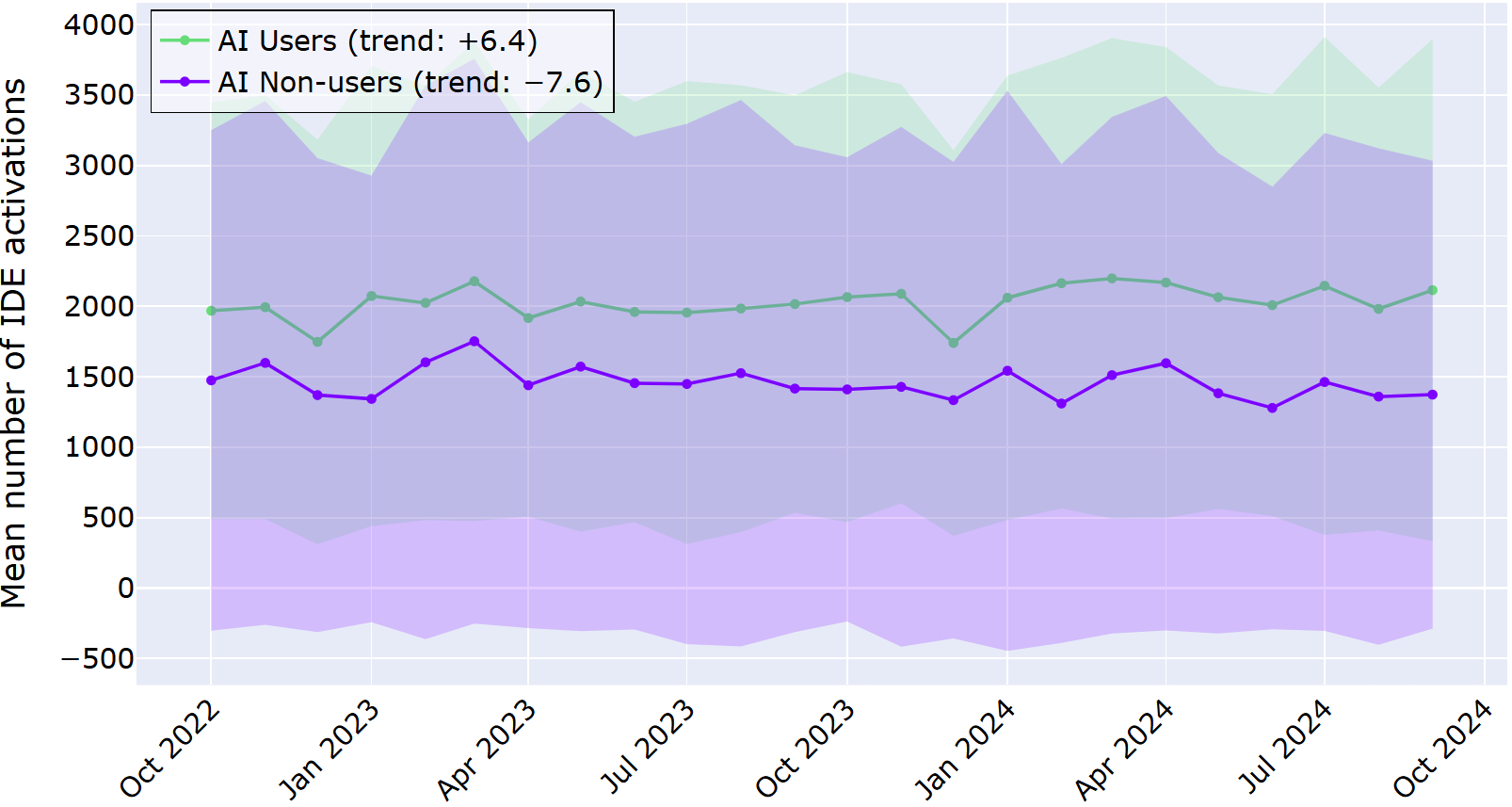}
    \caption{The mean \textit{Number of IDE activations} per device in the IDE logs, for different months. \textit{AI Users} show a statistically significant increasing trend of +6.4 activations a month, \textit{AI Non-users} --- a statistically significant decreasing trend of $-$7.6 activations a month. Shaded regions represent ±1 standard deviation from the monthly mean.}
    \label{fig:rq5:logs}
  \end{subfigure}

  \caption{Results for RQ5: Context switching.}
  \label{fig:rq5}
\end{figure}

Finally, the results for \textbf{context switching} are shown in Figure~\ref{fig:rq5}. 

\textbf{Survey and interviews} (Figure~\ref{fig:rq5:survey}). Here, we can see that 25.8\% of respondents indicated that their frequency of context switching slightly or significantly increased with the use of AI tools, 21\% --- that it decreased, and 53.2\% reported no change. P62 expressed a thought that highlights how the introduction of AI decreased context switching in their workflow:

\begin{quote}
    \textit{``I stopped switching contexts, saving a few seconds every time I would have googled something.''}
\end{quote}

\textbf{Logs} (Figure~\ref{fig:rq5:logs}). We can see that for \textit{AI users}, there is a slight statistically significant increase in the number of context switches with an estimated growth of approximately 6.4 switches per month. For \textit{AI non-users}, they had a slight statistically significant decrease of roughly 7.6 switches per month, as indicated by the negative interaction effect. This bi-directional trend highlights the difference in AI's influence on developers' behavior.

\observation{\textbf{Summary of RQ5}. Context switching demonstrated approximately an equal number of participants reporting it increasing and decreasing with the usage of AI tools. As for the number of window switches in the log data, it showed a slight increase for \textit{AI users}.}

\section{Discussion}
\label{section:discussion}

\begin{table*}[t]
\caption{An overview of the results of our study.}
\label{table:overview}
\begin{tabular}{cp{7cm}p{7cm}}
\toprule
\multicolumn{1}{l}{}      & \multicolumn{1}{c}{\textbf{Survey}} & \multicolumn{1}{c}{\textbf{IDE logs}} \\ \midrule
\textbf{Productivity}                                           &    \cellcolor{Green!70} Respondents are highly positive, 82.3\% say their productivity increased and 56.5\% say that their coding time decreased. &  \cellcolor{Green!70} The number of typed characters increases for both \textit{AI users} and \textit{AI non-users}, but grows much faster for \textit{AI users} (+587 characters vs +75 for \textit{AI non-users}).                                     \\ \midrule
\textbf{Code quality}                                         &   \cellcolor{YellowGreen!70} Respondents are relatively positive, 48.4\% say that code quality increased and 43.5\% see an increase in readability, with almost half not perceiving a change in both. &  \cellcolor{Goldenrod!70} There is no significant change for \textit{AI users} in terms of debugging sessions, while there is a slight negative trend of $-$0.46 sessions for \textit{AI non-users}.                                    \\ \midrule
\textbf{Code editing}                                          &   \cellcolor{YellowGreen!70} 43.5\% of respondents feel that the frequency of code editing increased, but 50\% see no change in it. &  \cellcolor{Green!70} The number of deletions increases for both \textit{AI users} and \textit{AI non-users}, but grows faster for \textit{AI users} (+102 deletions vs +7.6 for \textit{AI non-users}).                                     \\ \midrule
\textbf{Code reuse}        & \cellcolor{YellowGreen!70} Respondents are balanced in terms of using code from external sources, 37.1\% believe that the frequency of use of code from external sources increased, 19.4\% believe it decreased, and almost half see no change. &  \cellcolor{YellowGreen!70} Pasting from external sources demonstrates slight positive trends for both groups, the trend for \textit{AI users} (+1 paste per month) is higher than for \textit{AI non-users} (+0.4).                                                                                         \\ \midrule
\textbf{Context switching}                                       &   \cellcolor{Goldenrod!70} Respondents are split, 25.8\% indicated that frequency of context switching increased, while 21\% --- that it decreased, and 53.2\% reported no change. & \cellcolor{YellowGreen!70} There is a slight positive trend for IDE activations among \textit{AI users} (+6.4), and a slight negative trend for \textit{AI non-users} ($-$7.6). \\ \bottomrule                                      
\end{tabular}
\end{table*}

To understand the big picture impact of AI on developer workflows, we examine self-reported perceptions and behavioral telemetry. Table~\ref{table:overview} summarizes our findings across five key aspects.

\subsection{Relation of Aspects with Each Other}
Looking across the five workflow aspects, several inter-aspect patterns emerge that reveal how AI assistance influences not just isolated behaviors but broader development routines.

Perceived productivity gains among AI users are mirrored in telemetry through an increase in the amount of typing. Prior studies have similarly found that AI tools can accelerate code authoring~\cite{imai2022github, peng2023impact}. However, in logs, this increase in output is accompanied by a notable rise in code deletions. This suggests that productivity may come at a cost: developers produce more but also discard more, possibly due to iterative refinement of sometimes inefficient AI-generated suggestions~\cite{lertbanjongngam2022empirical}. Interestingly, no corresponding increase in debugging sessions was observed, which may indicate that errors are detected and addressed earlier in the workflow. The discrepancy between increased deletions and stable debugging aligns with observations from the literature that developers treat AI suggestions as starting points rather than final answers~\cite{weisz2022better}.

Such patterns echo earlier findings that developers often spend substantial time verifying and revising AI outputs~\cite{mozannar2024reading, sahoo2024ansible}. Our survey shows that more than a third of respondents perceive the code quality and need for code editing as growing, but half report no change in these metrics, which is expected due to the tendency for over-reliance by some users of AI tooling~\cite{prather2023s, schaffer2019can}. 

Code reuse and context switching also appear connected. In surveys, developers had no strong perception of change. Yet, in logs, we observe a slight but consistent increase in both pasting from external sources and activations reflecting window switches. While some literature anticipated a reduction in context switching due to in-IDE support~\cite{pandey2024transforming}, our data suggest that developers may still rely on external AI systems for validation or comparison. This hypothesis is supported by prior work documenting how developers use general-purpose tools like ChatGPT alongside IDE-integrated assistants~\cite{sergeyuk2024design}.

\subsection{Different Sources Tell Different Stories}
Behavioral telemetry and perceptual survey responses address the research questions from complementary angles: telemetry shows what developers do, surveys show what developers recognize. Taken together, our findings reveal that AI tools reshape developer workflows in ways that are observable at the event level but not always apparent in holistic retrospection. 

The telemetry data suggest that AI use is associated with increased activity volume in the IDE. Compared to \textit{AI non-users}, \textit{AI users} type and delete significantly more. In contrast, survey data portray a more reserved picture. Developers clearly report perceived productivity gains, but are more ambivalent across other aspects. For code quality, editing effort, reuse, and context switching, responses are often neutral, with roughly half of the participants reporting no observable change. These patterns echo prior work highlighting the disconnect between perceived and actual productivity~\cite{vaithilingam2022expectation, ziegler2022productivity}. While these patterns are consistent with AI-mediated workflow changes, we acknowledge that early adopters may represent a distinct population whose behaviors do not fully generalize to all developers. Our findings characterize developers who consistently adopted AI over a sustained period.

AI-assisted workflows appear to reconfigure where effort occurs, rather than how much effort is perceived. This subtle reconfiguration may partly explain the field's mixed findings on AI impact~\cite{song2024impact, liang2024large}, and highlights the importance of complementing telemetry with perceptual data when evaluating developer-facing AI tools. Both perspectives are valid and complementary. Understanding AI's real-world impact requires examining both.

Overall, the combined picture points to a reframing of AI's role in development workflows: not as a direct substitute for effort or expertise, but as a modifier of rhythm, density, and focus of interaction. Rather than producing sweeping transformations visible to the user, AI tools appear to reshape the granularity and structure of everyday tasks. Recognizing this distinction is essential for both empirical evaluation and the design of future AI tools.

\subsection{Durability of Findings in Evolving AI Landscapes}

Our study captures developer workflows during a critical transition period when AI assistance primarily operated through code completion and in-chat suggestions. The landscape has continued to evolve rapidly. Current agentic coding systems that autonomously generate larger code units are continuing to reshape development practice~\cite{yang2024swe, cognition2024devin}. This invites consideration of how our findings may evolve alongside advancing AI capabilities. 

We argue that several core insights are durable. First, the gap between perception and behavior that we identify reflects fundamental human cognitive limitations in retrospective self-assessment~\cite{kahneman2011thinking}, not properties of specific AI technologies. Whether code is generated via tab completion or autonomous agents, developers may similarly misjudge the extent to which their workflows have transformed. Second, the pattern of increased output coupled with increased deletion suggests a general dynamic of AI-assisted work: greater volume requires greater curation~\cite{mozannar2024reading}. As AI generates more code, the editing and validation burden likely intensifies rather than disappears. Even as agents assume more authoring responsibility, developers must allocate their attention to specification, verification, and integration tasks.

Our work provides an empirical baseline for understanding how AI adoption reshaped workflows, enabling future research to measure whether subsequent AI paradigms amplify, reverse, or fundamentally alter these patterns. The complementary telemetry-survey methodology we demonstrate remains applicable for studying newer AI systems, even as the specific metrics must adapt to changing interaction modalities.

\subsection{Design Implications for Future AI-Augmented IDEs} Our findings suggest that existing AI assistants are designed primarily for localized support (\textit{e.g.}, completing code) but lack awareness of broader workflow impacts. As development environments evolve toward more agentic systems and AI-native platforms, these limitations may become more pronounced or take different forms. To better support developers, future AI tools should not only generate code but also support context continuity, anticipate multi-step task flows, and provide visibility into where and how assistance is influencing developer behavior.

Furthermore, tools should support reflective practices—surfacing insights about the developer’s workflow evolution, rather than remaining silent partners. This becomes increasingly critical as AI assumes more autonomous roles: when agents generate substantial code, developers need explicit signals about changes in their verification effort, integration patterns, and overall workflow. For instance, telemetry-informed feedback on changes in reuse patterns or editing intensity could help developers regain situational awareness and avoid overreliance.

\section{Threats to Validity}
\label{section:threats}
\paragraph{Construct validity.}
Grounded in prior work, we operationalized five workflow dimensions using in-IDE telemetry and survey responses. We deliberately included both behavioral and perceptual data, anticipating that they may represent different constructs. While telemetry ensures consistency and avoids recall bias, capturing fine-grained, moment-to-moment developer actions, it does not capture developer intent or task context. Conversely, survey data reflects holistic, retrospective perception but not actual behavior. To address these limitations, we complement data sources. In our case, these sources sometimes point in different directions. We do not treat this as a weakness, but as evidence that each source highlights a different facet of AI-assisted workflows. To interpret our findings, readers should understand that behavioral telemetry and survey responses are not interchangeable measures of the same construct, but complementary windows into developer experience.

The study is based on telemetry from a single commercial IDE provider, JetBrains, and uses integration with a specific proprietary JetBrains AI assistant to define AI usage. Developers using other AI tools or AI tools outside the IDE may be misclassified as non-users, potentially underestimating true adoption. However, given the assistant's full integration and free trial availability, we assume that developers interested in using AI within the IDE would have at least tried the built-in assistant. Those who never activated it over two years, even once, are likely less inclined to adopt in-IDE AI support. Given that this was the only criterion for sampling the two separate groups, numerous statistically significant results indicate that the change is measurable, even if it does not fully generalize to users of all other AI tools.

\paragraph{Internal validity.}
Our data confirms that \textit{AI users} are generally more active in the IDE than \textit{AI non-users}, as evidenced by the fact that for most actions, \textit{AI users} do more of them. It is likely that they differ not only in AI usage but also in baseline motivation, experience, or task profiles --- all of which may influence AI adoption and observed behavior. However, our focus is on behavioral trends within each group over time, rather than cross-group causality. As such, we treat the distinction between \textit{AI users} and \textit{AI non-users} as descriptive rather than explanatory. For instance, when we observe that \textit{AI users} show steeper increases in typed characters and deletions compared to \textit{AI non-users}, this pattern could reflect AI's direct influence on coding intensity. Alternatively, it may indicate that early adopters maintain elevated activity levels regardless, and that AI integration occurred within an already active population. These interpretations cannot be fully disentangled without experimental assignment to conditions. The longitudinal nature of the analysis of trends supports the validity of comparisons within each group, while acknowledging that self-selection effects may shape between-group differences. Readers should interpret cross-group differences as associations reflecting both AI adoption and the unmeasured characteristics that predict adoption.

\paragraph{External validity.}
While the studied AI assistant offers general-purpose coding support comparable to other AI-based tools, we acknowledge that our findings may not generalize to all development environments or interface paradigms. At the same time, our data includes users of different IDEs, different programming languages, and different ecosystems. Moreover, the workflow dimensions we examine --- productivity, code quality, code editing, code reuse, and context switching --- are not tool-specific and remain relevant across a wide range of IDEs and development contexts. This supports the partial generalizability of our results beyond the specific assistant and platform studied. 

\section{Conclusion}
\label{section:conclusion}
We conducted a mixed-method study of AI adoption in IDEs, combining two-year longitudinal fine-grained telemetry from 800 professional developers, capturing over 151 million interaction events, with survey responses from 62 participants. Our findings reveal that AI-assisted development workflows evolve in subtle but measurable ways over time. While \textit{AI users} consistently report increased productivity and minimal changes in other dimensions, fine-grained telemetry traces show significant increases in writing and editing activity, as well as a rising trend in external code integration, and context switching. These results show that AI redistributes and reshapes development work, frequently in ways that elude developers’ conscious awareness.

As AI becomes further integrated into the software development lifecycle, our results underscore the need for tooling and evaluation methods that account for both visible and more subtle, often overlooked shifts in workflow. Future research and tool design should focus not only on output metrics but also on workflow granularity, cognitive load, and user awareness to ensure that AI not only supports productivity but also aligns with and enhances core software engineering practices, rather than unintentionally reshaping them in opaque ways.

\section*{Data Availability}

Supplementary materials, including the survey questionnaire and anonymized responses, interview script, and complete statistical analysis outputs, are publicly available in our supplementary materials repository~\cite{artifacts}.

Raw IDE telemetry logs cannot be released due to confidentiality agreements with our industry partner. All aggregated data and statistical outputs necessary to verify our findings are provided in the supplementary materials.

Researchers interested in accessing additional aggregated data or replicating our analysis methodology may contact the authors.

\bibliographystyle{ACM-Reference-Format}
\balance
\bibliography{refs}

@inproceedings{minelli2015know,
  title={I know what you did last summer-an investigation of how developers spend their time},
  author={Minelli, Roberto and Mocci, Andrea and Lanza, Michele},
  booktitle={2015 IEEE 23rd international conference on program comprehension},
  pages={25--35},
  year={2015},
  organization={IEEE}
}

@inproceedings{amann2016study,
  title={A study of {Visual Studio} usage in practice},
  author={Amann, Sven and Proksch, Sebastian and Nadi, Sarah and Mezini, Mira},
  booktitle={2016 IEEE 23rd International Conference on Software Analysis, Evolution, and Reengineering (SANER)},
  volume={1},
  pages={124--134},
  year={2016},
  organization={IEEE}
}

@article{astromskis2017patterns,
  title={Patterns of developers behaviour: A 1000-hour industrial study},
  author={Astromskis, Saulius and Bavota, Gabriele and Janes, Andrea and Russo, Barbara and Di Penta, Massimiliano},
  journal={Journal of Systems and Software},
  volume={132},
  pages={85--97},
  year={2017},
  publisher={Elsevier}
}

@inproceedings{damevski2016interactive,
  title={Interactive exploration of developer interaction traces using a hidden Markov model},
  author={Damevski, Kostadin and Chen, Hui and Shepherd, David and Pollock, Lori},
  booktitle={Proceedings of the 13th International Conference on Mining Software Repositories},
  pages={126--136},
  year={2016}
}

@inproceedings{coutinho2024role,
  title={The role of generative AI in software development productivity: A pilot case study},
  author={Coutinho, Mariana and Marques, Lorena and Santos, Anderson and Dahia, Marcio and Fran{\c{c}}a, Cesar and de Souza Santos, Ronnie},
  booktitle={Proceedings of the 1st ACM International Conference on AI-Powered Software},
  pages={131--138},
  year={2024}
}

@inproceedings{imai2022github,
  title={Is GitHub Copilot a substitute for human pair-programming? An empirical study},
  author={Imai, Saki},
  booktitle={Proceedings of the ACM/IEEE 44th International Conference on Software Engineering: Companion Proceedings},
  pages={319--321},
  year={2022}
}

@article{dakhel2023github,
  title={GitHub Copilot AI pair programmer: Asset or liability?},
  author={Dakhel, Arghavan Moradi and Majdinasab, Vahid and Nikanjam, Amin and Khomh, Foutse and Desmarais, Michel C and Jiang, Zhen Ming Jack},
  journal={Journal of Systems and Software},
  volume={203},
  pages={111734},
  year={2023},
  publisher={Elsevier}
}

@inproceedings{ross2023programmer,
  title={The programmer’s assistant: Conversational interaction with a large language model for software development},
  author={Ross, Steven I and Martinez, Fernando and Houde, Stephanie and Muller, Michael and Weisz, Justin D},
  booktitle={Proceedings of the 28th International Conference on Intelligent User Interfaces},
  pages={491--514},
  year={2023}
}

@inproceedings{vaithilingam2022expectation,
  title={Expectation vs. experience: Evaluating the usability of code generation tools powered by large language models},
  author={Vaithilingam, Priyan and Zhang, Tianyi and Glassman, Elena L},
  booktitle={Chi conference on human factors in computing systems extended abstracts},
  pages={1--7},
  year={2022}
}

@inproceedings{weisz2022better,
  title={Better together? An evaluation of AI-supported code translation},
  author={Weisz, Justin D and Muller, Michael and Ross, Steven I and Martinez, Fernando and Houde, Stephanie and Agarwal, Mayank and Talamadupula, Kartik and Richards, John T},
  booktitle={Proceedings of the 27th International Conference on Intelligent User Interfaces},
  pages={369--391},
  year={2022}
}

@inproceedings{ziegler2022productivity,
  title={Productivity assessment of neural code completion},
  author={Ziegler, Albert and Kalliamvakou, Eirini and Li, X Alice and Rice, Andrew and Rifkin, Devon and Simister, Shawn and Sittampalam, Ganesh and Aftandilian, Edward},
  booktitle={Proceedings of the 6th ACM SIGPLAN International Symposium on Machine Programming},
  pages={21--29},
  year={2022}
}

@article{song2024impact,
  title={The impact of generative AI on collaborative open-source software development: Evidence from GitHub Copilot},
  author={Song, Fangchen and Agarwal, Ashish and Wen, Wen},
  journal={arXiv preprint arXiv:2410.02091},
  year={2024}
}

@article{weber2024significant,
  title={Significant productivity gains through programming with large language models},
  author={Weber, Thomas and Brandmaier, Maximilian and Schmidt, Albrecht and Mayer, Sven},
  journal={Proceedings of the ACM on Human-Computer Interaction},
  volume={8},
  number={EICS},
  pages={1--29},
  year={2024},
  publisher={ACM New York, NY, USA}
}

@inproceedings{sahoo2024ansible,
  title={Ansible Lightspeed: A code generation service for IT automation},
  author={Sahoo, Priyam and Pujar, Saurabh and Nalawade, Ganesh and Genhardt, Richard and Mandel, Louis and Buratti, Luca},
  booktitle={Proceedings of the 39th IEEE/ACM International Conference on Automated Software Engineering},
  pages={2148--2158},
  year={2024}
}

@inproceedings{mozannar2024reading,
  title={Reading between the lines: Modeling user behavior and costs in AI-assisted programming},
  author={Mozannar, Hussein and Bansal, Gagan and Fourney, Adam and Horvitz, Eric},
  booktitle={Proceedings of the CHI Conference on Human Factors in Computing Systems},
  pages={1--16},
  year={2024}
}

@article{prather2023s,
  title={“It’s weird that it knows what I want”: Usability and interactions with Copilot for novice programmers},
  author={Prather, James and Reeves, Brent N and Denny, Paul and Becker, Brett A and Leinonen, Juho and Luxton-Reilly, Andrew and Powell, Garrett and Finnie-Ansley, James and Santos, Eddie Antonio},
  journal={ACM Transactions on Computer-Human Interaction},
  volume={31},
  number={1},
  pages={1--31},
  year={2023},
  publisher={ACM New York, NY}
}

@inproceedings{schaffer2019can,
  title={I can do better than your AI: expertise and explanations},
  author={Schaffer, James and O'Donovan, John and Michaelis, James and Raglin, Adrienne and H{\"o}llerer, Tobias},
  booktitle={Proceedings of the 24th International Conference on Intelligent User Interfaces},
  pages={240--251},
  year={2019}
}

@inproceedings{lertbanjongngam2022empirical,
  title={An empirical evaluation of competitive programming AI: A case study of Alphacode},
  author={Lertbanjongngam, Sila and Chinthanet, Bodin and Ishio, Takashi and Kula, Raula Gaikovina and Leelaprute, Pattara and Manaskasemsak, Bundit and Rungsawang, Arnon and Matsumoto, Kenichi},
  booktitle={2022 IEEE 16th International Workshop on Software Clones (IWSC)},
  pages={10--15},
  year={2022},
  organization={IEEE}
}

@article{sergeyuk2024design,
  title={The Design Space of in-IDE Human-AI Experience},
  author={Sergeyuk, Agnia and Koshchenko, Ekaterina and Zakharov, Ilya and Bryksin, Timofey and Izadi, Maliheh},
  journal={arXiv preprint arXiv:2410.08676},
  year={2024}
}

@article{Chen2021,
  title={Evaluating large language models trained on code},
  author={Chen, Mark and Tworek, Jerry and Jun, Heewoo and Yuan, Qiming and Pinto, Henrique Ponde De Oliveira and Kaplan, Jared and Edwards, Harri and Burda, Yuri and Joseph, Nicholas and Brockman, Greg and others},
  journal={arXiv preprint arXiv:2107.03374},
  year={2021}
}

@misc{iccesomar,
  title = {{ICC/ESOMAR} codes and guidelines},
  howpublished = {\url{https://esomar.org/codes-and-guidelines}},
  note = {Accessed: 2025-07-14}
}

@misc{openai2022chatgpt,
  author    = {OpenAI},
  title     = {ChatGPT: Mar 14 version},
  year      = {2022},
  url       = {https://openai.com/blog/chatgpt},
  note         = {Accessed: 2025-07-14},
}

@article{del2023large,
  title={Are large language models a threat to digital public goods? Evidence from activity on Stack Overflow},
  author={del Rio-Chanona, Maria and Laurentsyeva, Nadzeya and Wachs, Johannes},
  journal={arXiv preprint arXiv:2307.07367},
  year={2023}
}

@article{pandey2024transforming,
  title={Transforming software development: Evaluating the efficiency and challenges of GitHub Copilot in real-world projects},
  author={Pandey, Ruchika and Singh, Prabhat and Wei, Raymond and Shankar, Shaila},
  journal={arXiv preprint arXiv:2406.17910},
  year={2024}
}

@article{vaillant2024developers,
  title={Developers' perceptions on the impact of ChatGPT in software development: A survey},
  author={Vaillant, Thiago S and de Almeida, Felipe Deveza and Neto, Paulo Anselmo and Gao, Cuiyun and Bosch, Jan and de Almeida, Eduardo Santana},
  journal={arXiv preprint arXiv:2405.12195},
  year={2024}
}

@article{peng2023impact,
  title={The impact of AI on developer productivity: Evidence from GitHub Copilot},
  author={Peng, Sida and Kalliamvakou, Eirini and Cihon, Peter and Demirer, Mert},
  journal={arXiv preprint arXiv:2302.06590},
  year={2023}
}

@inproceedings{liang2024large,
  title={A large-scale survey on the usability of AI programming assistants: Successes and challenges},
  author={Liang, Jenny T and Yang, Chenyang and Myers, Brad A},
  booktitle={Proceedings of the 46th IEEE/ACM international conference on software engineering},
  pages={1--13},
  year={2024}
}

@article{forsgren2021space,
  title={The SPACE of developer productivity: There's more to it than you think},
  author={Forsgren, Nicole and Storey, Margaret-Anne and Maddila, Chandra and Zimmermann, Thomas and Houck, Brian and Butler, Jenna},
  journal={Queue},
  volume={19},
  number={1},
  pages={20--48},
  year={2021},
  publisher={ACM New York, NY, USA}
}

@article{murphy2019predicts,
  title={What predicts software developers’ productivity?},
  author={Murphy-Hill, Emerson and Jaspan, Ciera and Sadowski, Caitlin and Shepherd, David and Phillips, Michael and Winter, Collin and Knight, Andrea and Smith, Edward and Jorde, Matthew},
  journal={IEEE Transactions on Software Engineering},
  volume={47},
  number={3},
  pages={582--594},
  year={2019},
  publisher={IEEE}
}

@article{tabarsi2025llms,
  title={LLMs' reshaping of people, processes, products, and society in software development: A comprehensive exploration with early adopters},
  author={Tabarsi, Benyamin and Reichert, Heidi and Limke, Ally and Kuttal, Sandeep and Barnes, Tiffany},
  journal={arXiv preprint arXiv:2503.05012},
  year={2025}
}

@article{sergeyuk2025using,
  title={Using AI-based coding assistants in practice: State of affairs, perceptions, and ways forward},
  author={Sergeyuk, Agnia and Golubev, Yaroslav and Bryksin, Timofey and Ahmed, Iftekhar},
  journal={Information and Software Technology},
  volume={178},
  pages={107610},
  year={2025},
  publisher={Elsevier}
}

@misc{JetBrains2024DevEcosystem,
  author       = {{JetBrains}},
  title        = {The State of Developer Ecosystem 2024},
  howpublished = {\url{https://www.jetbrains.com/lp/devecosystem-2024/}},
  note         = {Accessed 2025-07-14},
  year         = {2024}
}

@misc{StackOverflow2024Survey,
  author       = {{Stack Overflow}},
  title        = {Stack Overflow developer survey 2024},
  howpublished = {\url{https://survey.stackoverflow.co/2024/ai}},
  note         = {Accessed 2025-07-14},
  year         = {2024}
}

@techreport{dell2025cybernetic,
  title={The cybernetic teammate: A field experiment on generative AI reshaping teamwork and expertise},
  author={Dell'Acqua, Fabrizio and Ayoubi, Charles and Lifshitz, Hila and Sadun, Raffaella and Mollick, Ethan and Mollick, Lilach and Han, Yi and Goldman, Jeff and Nair, Hari and Taub, Stewart and others},
  year={2025},
  institution={National Bureau of Economic Research}
}

@article{beckermeasuring,
  title={Measuring the impact of early-2025 AI on experienced open-source developer productivity},
  author={Becker, Joel and Rush, Nate and Barnes, Elizabeth and Rein, David},
  journal={arXiv preprint arXiv:2507.09089},
  year={2025}
}

@inproceedings{devanbu2016belief,
  title={Belief \& evidence in empirical software engineering},
  author={Devanbu, Prem and Zimmermann, Thomas and Bird, Christian},
  booktitle={Proceedings of the 38th international conference on software engineering},
  pages={108--119},
  year={2016}
}

@misc{jetbrains_ai,
  title        = {JetBrains AI Assistant},
  author       = {{JetBrains}},
  note         = {Accessed: 2025-07-14},
  url          = {https://plugins.jetbrains.com/plugin/22282-jetbrains-ai-assistant}
}

@misc{github_copilot,
  title        = {GitHub Copilot},
  author       = {{GitHub and OpenAI}},
  note         = {Accessed: 2025-07-14},
  url          = {https://github.com/features/copilot/}
}

@techreport{harding2025aicodequality,
  author       = {William Harding},
  title        = {AI Copilot code quality: Evaluating 2024's increased defect rate via code quality metrics},
  institution  = {Alloy.dev Research},
  year         = {2025},
  url          = {https://www.gitclear.com/ai_assistant_code_quality_2025_research},
  note         = {Accessed: 2025-07-14}
}

@manual{scipy_kstwo,
  title        = {scipy.stats.kstwo — SciPy v1.13.0 Manual},
  author       = {{SciPy}},
  year         = {2024},
  url          = {https://docs.scipy.org/doc/scipy/reference/generated/scipy.stats.kstwo.html},
  note         = {Accessed: 2025-07-14}
}

@manual{scipy_bartlett,
  title        = {scipy.stats.bartlett — SciPy v1.13.0 Manual},
  author       = {{SciPy}},
  year         = {2024},
  url          = {https://docs.scipy.org/doc/scipy/reference/generated/scipy.stats.bartlett.html},
  note         = {Accessed: 2025-07-14}
}

@manual{statsmodels_mixedlm,
  title        = {Mixed Linear Model — statsmodels v0.14.0 Manual},
  author       = {{statsmodels}},
  year         = {2024},
  url          = {https://www.statsmodels.org/stable/mixed_linear.html},
  note         = {Accessed: 2025-07-14}
}

@misc{artifacts,
  title = {Supplementary materials},
  year = {2025},
  author = {Sergeyuk, Agnia and Huang, Eric and Karaeva, Dariia and Serova, Anastasiia and Golubev, Yaroslav and Ahmed, Iftekhar},
  howpublished = {\url{https://doi.org/10.5281/zenodo.18220544}},
  note = {Accessed: 2026-01-12}
}

@article{yang2024swe,
  title={Swe-agent: Agent-computer interfaces enable automated software engineering},
  author={Yang, John and Jimenez, Carlos E and Wettig, Alexander and Lieret, Kilian and Yao, Shunyu and Narasimhan, Karthik and Press, Ofir},
  journal={Advances in Neural Information Processing Systems},
  volume={37},
  pages={50528--50652},
  year={2024}
}

@misc{cognition2024devin,
  title={Devin: AI Software Engineer},
  author={{Cognition AI}},
  year={2024},
  howpublished={\url{https://www.cognition.ai/blog/introducing-devin}},
  note={Accessed: 2025-01-15}
}

@book{kahneman2011thinking,
  title={Thinking, Fast and Slow},
  author={Kahneman, Daniel},
  year={2011},
  publisher={Farrar, Straus and Giroux}
}

\end{document}